\documentclass[fleqn,usenatbib]{mnras}

\usepackage{newtxtext,newtxmath}

\usepackage[T1]{fontenc}

\DeclareRobustCommand{\VAN}[3]{#2}
\let\VANthebibliography\thebibliography
\def\thebibliography{\DeclareRobustCommand{\VAN}[3]{##3}\VANthebibliography}

\usepackage{graphicx}	
\usepackage{amsmath}	
\usepackage{subcaption}

\newcommand{\lw}{\textit{Lightweaver}}
\newcommand{\ch}{\textsc{CHIANTI}}

\title[Improving the atomic modelling in RT]{Improving the atomic modelling for solar UV radiative transfer calculations}

\author[R.P. Dufresne et al.]{
R.P. Dufresne,$^{1}$\thanks{E-mail: rpd21@cam.ac.uk}
G. Del Zanna,$^{1,2}$
and C.M.J. Osborne$^{3}$
\\
$^{1}$DAMTP, University of Cambridge, Wilberforce Road, Cambridge CB3 0WA, UK\\
$^{2}$School of Physics and Astronomy, University of Leicester, Leicester LE1 7RH, UK\\
$^{3}$SUPA School of Physics and Astronomy, University of Glasgow, Glasgow G12 8QQ, UK\\
}

\date{Accepted XXX. Received YYY; in original form ZZZ}

\pubyear{\the\year{}}

\begin{document}
\label{firstpage}
\pagerange{\pageref{firstpage}--\pageref{lastpage}}
\maketitle

\begin{abstract}
Radiative transfer calculations have been produced over the years for many lines and continua in the UV wavelength range of solar and cool stellar atmospheres for a variety of conditions. Despite significant improvements in computing power and availability of atomic data over time, atomic models are often still limited in size and rely on approximations for data. There have also been inconsistencies in the way photo-ionisation and radiative recombination have been treated. Here, we incorporate into the Lightweaver radiative transfer code new data and updated modelling of atomic processes for the low charge states of C, Si and S. Data are taken from the CHIANTI database and other widely-available sources for the relevant elements. We show the significant impact this has on the UV continua in the 1100-1700\,\AA\ region, especially for Si. The results are in much better agreement with averaged, quiet Sun observations, and remove the need to invoke ``missing opacity'' to resolve discrepancies. The present treatment has important implications for radiative transfer calculations and the model atmospheres used as inputs.

\end{abstract}




\section{Introduction}
\label{sec:intro}


Stellar chromospheres and lower transition regions emit primarily in the ultra-violet (UV), in the 
form of spectral lines and continua. In the early days of UV observations, very simplified atomic models had to be used in radiative transfer (RT) codes when investigating such spectral features which are affected by opacity. For instance, \citet{finn1969} investigated Lyman-$\alpha$ and -$\beta$ emission using a hydrogen model with three levels. Analytical expressions for atomic processes were especially useful because of limitations in computing resources. Another example includes them utilising the widely-used \citet{vanregemorter1962} approximation for electron impact excitation (EIE) in a two level model studying an \ion{Al}{i} line in the solar atmosphere \citep{finn1974}.

While these were often the only available sources of data at the time, use of such expressions continued. Moving forward almost three decades and the RH code \citep{uitenbroek2001}, for example, made extensive use of the \citet{seaton1962} impact parameter method for EIE. This, like the \citet{vanregemorter1962} approach, is a very rough approximation and unsuitable for neutrals and low charge states. Another commonly used approximation is the general formula of \citet{burgess1965dr} for dielectronic recombination (DR), which is a significant process in the solar atmosphere \citep{burgess1964dr}. The formula is only suitable for high temperature settings like the solar corona and does not take account of effects that are important at lower temperatures \citep{storey1981}. However, the \citet{burgess1965dr} formula is routinely found in codes studying this region \citep[such as][]{olluri2013bifrost}, either directly or indirectly through rates produced by others \citep[][for example]{arnaud1985,shull1982}. 

Even where \textit{ab initio} data is used, it is often unsuitable for the setting in which it is being used. The photo-ionisation (PI) cross sections produced by the Opacity Project \citep[OP,][]{seaton1995op} were state-of-the-art R-Matrix calculations that became the reference for many later calculations. The data are used in RT codes to calculate the reverse reaction, radiative recombination (RR), but it can neglect important contributions to the rates, as will be discussed later. Much of the charge transfer data used in codes comes from compilations like \citet{arnaud1985,kingdon1996}. This was calculated using assumptions that means it is unsuitable for the temperatures in lower solar and stellar atmospheres \citep{bates1962ct}. This data has often been used \citep[such as in][]{avrett2008}, even though there were many works which carried out more accurate calculations in the meantime \citep[including][for example]{stancil1997si5he,mroczkowski2003}. 

In the intervening years since the earliest RT calculations, there have been many improvements in radiative transfer methods. The earlier codes assumed static 1D atmospheres, with various corrections to take into account dynamics (e.g. line broadenings). The same codes were also used to obtain average 1D atmospheric models by comparison with observations. A widely-used code, PANDORA, written by R. Loeser under the guidance of E. Avrett, formed a basis used by many codes \citep[see][and following papers]{vernazza_etal:1981,fal:1990}. MULTI is another widely-used code \citep{multi:1986}. The important methods developed by \cite{rybicki_hummer:1992} were included in RH. Within astrophysics many RT codes have been written, but there are too many to be listed here. As an example, an older one is TLUSTY \citep[see][and following papers]{hubeny:1995}, whilst a recent one is MCFOST \citep{2021A&A...647A..27T}. For a detailed review of non-LTE methods (that is, for departures from local thermodynamic equilibrium) and codes, refer to \citet{hubeny2015book}. Most modern radiative transfer codes use the multilevel accelerated lambda iteration (MALI) method described in two papers by \citet{rybicki_hummer:1991, rybicki_hummer:1992}. An additional secondary linear Newton-Raphson iteration coupled to MALI to perform charge conservation was described for hydrogen by \citet{heinzel1995} and extended to multiple atomic species within the framework of full-preconditioning MALI by \citet{osborne2021lw}.

There have been further significant improvements in RT calculations in recent decades by, for example, coupling RT with time-dependent ionisation and MHD \citep[see the review by][]{2020LRSP...17....3L}, and including partial redistribution in 3D non-LTE calculations \citep[see][]{2017A&A...597A..46s}. These more advanced RT models are clearly necessary to gain a better understanding of chromospheres, and it is understandable that reduced models may be required for such complex studies. However, the feedback of atomic processes is a fundamental part of the modelling, and yet inspection of the atomic models used by these codes makes it appear that many of the models may be inadequate for the task.

The aim of this paper is to provide an assessment of the impact better quality atomic data and modelling have on RT calculations. To this end, we utilise the publicly-available, \lw\ \citep{osborne2021lw} RT code. This code has been benchmarked against several of the above-mentioned codes and uses similar approximations and atomic data as the previous ones. We import as much atomic data as possible from the \ch\ database \citep{dere1997,dufresne2024v11}. Beyond that, we obtain data that is readily available from other sources, and make new calculations to ensure a consistent treatment only where necessary. Although we use a static 1D model of the quiet Sun, such models are still widely-used and are useful for a variety of studies. Here, we focus on the UV region of the solar atmosphere (limiting our discussion to the wavelength range 1100-1700\,\AA) to illustrate the main effects of improving the atomic models when calculating the continuum.

The paper is structured as follows: Section\;\ref{sec:methods} describes the methods, summarizes the old and new atomic models, and briefly describes the adopted model atmosphere and quiet Sun observations used for the comparisons. Section\;\ref{sec:results} describes the main results, Sect.\;\ref{sec:discussion} provides a discussion on the implications of the present results and potential future improvements. Section\;\ref{sec:concl} summarises the main findings.

\section{Methods}
\label{sec:methods}

\subsection{The \lw\ radiative transfer code}
\label{sec:rt}

In this work we use the \lw{} non-LTE radiative transfer framework \citep{osborne2021lw}.
\lw{} employs similar methods to RH, namely the full-preconditioning method of \citet{rybicki_hummer:1992}, with support for partial frequency redistribution in either the angle-averaged formulation, or the hybrid method of \citet{leenaarts_etal:2012}.
Here, we utilise the third-order Bézier formal solver of \citet{delacruzrodriguez_piskunov:2013}.

\lw{} differs from other radiative transfer codes in its design as a framework, rather than a fixed-function code: through Python the user can make deep modifications and investigations into intermediate values that are often hidden in other tools, or construct an entirely new tool for a specific purpose.
It has been validated against RH by \citet{osborne2021lw}, SNAPI \citep{milic_vannoort:2018} by \citet{osborne_thesis:2021}, and RADYN \citep{carlsson_stein:1992, allred_etal:2015} by \citet{osborne_etal:2021}.

\subsection{The older atomic models}
\label{sec:lwatoms}

Various atomic models, as made available in \lw, have been used here as a reference against which to compare the new atomic models. These older models were all taken from the RH code. The ones used here are denoted in \lw\ as \texttt{H\_6\_atom}, \texttt{He\_9\_atom}, \texttt{C\_atom}, \texttt{N\_atom}, \texttt{OI\_fine\_atom},  \texttt{MgI\_bMgII\_atom}, \texttt{Al\_atom}, \texttt{Si\_atom}, \texttt{S\_atom}, \texttt{Fe\_43\_61\_atom} and \texttt{NiI\_atom}. All elements were run as detailed active atoms, that is, solved in non-LTE, except for N, S and Ni. N was solved using LTE populations because it does not make noticeable contributions to the continuum in the wavelength region of interest; the old atomic model for S only has continua and no collisions and radiative decays; while for Ni it was not clear how much it contributed to the continuum in LTE because of overlaps in its edges with other elements, plus the Ni model would not converge with the present setup. The above set of atomic models are collectively referred to as the base set of models throughout the rest of this paper. Complete frequency redistribution was used throughout the work in all cases except for the Lyman-$\alpha$ and -$\beta$ lines, which were treated with partial frequency redistribution. We tested the effect of Thomson scattering and Rayleigh scattering off H, He and H$_2$, and found it altered the continuum intensities by 1-2\% at the longer wavelengths of interest here and not at all at the shorter wavelengths. Consequently, we did not include these processes during the calculations.

\subsection{The new atomic models}
\label{sec:atoms}

The radiative transfer calculation featuring the new atomic models will be compared to the base set of models in order to highlight the differences the new models make to the synthetic spectra and other parameters of the solar atmosphere. Everything is kept the same between the two runs except that changes are made to the atomic models of C, Si and S. Completely new models were constructed for these elements, which will now be described.

\subsubsection{Photo-ionisation}

The photo-ionisation (PI) cross sections produced by the Opacity Project were primarily designed for Rosseland mean opacities in stellar envelopes and, in such cases, only the total cross sections from each atomic level are of importance. Using the OP data in solar and stellar atmospheres, however, can present several problems. The main problem with using total PI cross sections arises because RT codes generally use them to calculate the reverse RR rates. This underestimates the RR rates, as described in the next section. Another potential problem is that using PI cross sections totalled over final states and posted to one level in the next higher charge state can alter the level populations within the final state. This is an issue more in low-density plasma because ionisation to individual levels is not properly accounted for in those cases and transitions amongst bound states do not redistribute the imbalanced populations in such conditions.

To ensure PI and RR are treated correctly, photo-ionisation cross sections resolved by initial and final level should be used. Such cross sections were calculated by \citet{badnell2006} for all H- to Mg-like ions of H to Zn, and have been made available on the UK Atomic Processes for Astrophysical Plasmas (APAP) website\footnote{http://www.apap-network.org}. These cross sections were used here for applicable ions in the new models. For the relevant ions not covered by that work (\ion{Si}{i-ii} and \ion{S}{i-iii}), level-resolved cross sections were calculated using {\sc Autostructure} \citep[AS][]{badnell2011}, the same program used by \citet{badnell2006}.

\begin{figure}
\centering
\begin{subfigure}{0.45\textwidth}
\includegraphics[width=\textwidth]{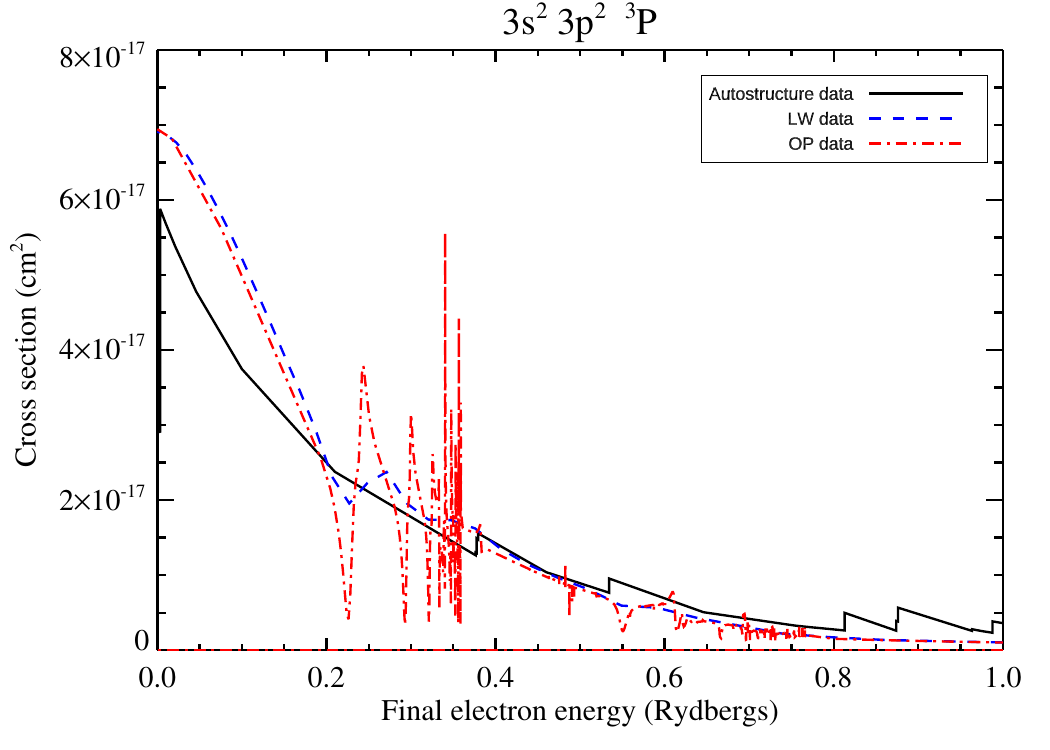}
\caption{Total photo-ionisation cross sections for the ground state of Si$^0$.}
\label{fig:si1pics}
\end{subfigure}
\hfill
\begin{subfigure}{0.45\textwidth}
\includegraphics[width=\textwidth]{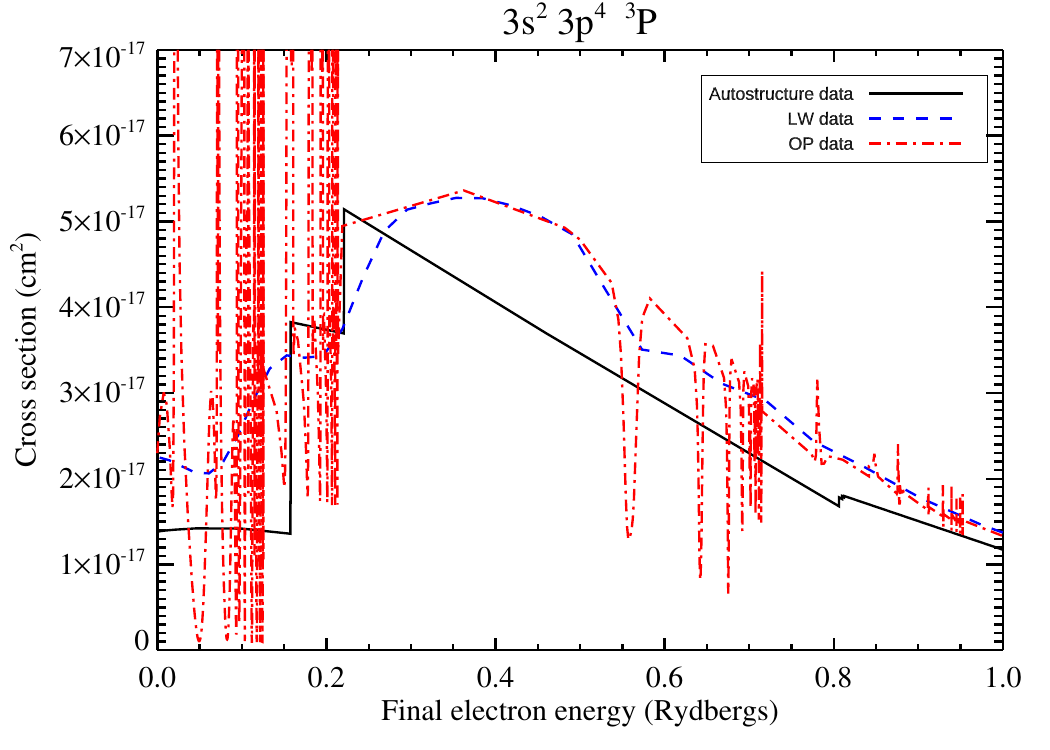}
\caption{Total photo-ionisation cross sections for the ground state of S$^0$.}
\label{fig:s1pics}
\end{subfigure}

\caption{Comparison of total photo-ionisation cross sections from the widely-used OP data, the smoothed cross sections used in \lw, and the level-resolved calculations from \textsc{Autostructure} used for the present work.}
\label{fig:pics}
\end{figure}

The results of these calculations for the ground states of \ion{Si}{i} and \ion{S}{i} are shown in Fig.\ref{fig:pics}. The level-resolved cross sections are totalled over all final states in the illustration to compare with the OP results. AS calculates all continuum processes in the distorted-wave approximation, which does not include resonances caused by photo-excitation into auto-ionising states, followed by decay into the continuum. Such resonances are included in the OP calculations, but are typically excluded or under-sampled in RT calculations, for example as in \lw\ and RH. This means the important comparison here is that the distorted-wave calculation matches the background \textit{R}-matrix cross section. This is clearly the case with the new calculations for Si$^0$ and S$^0$ shown here. In fact, the distorted wave cross sections make it possible to see where the background cross section is in the S$^0$ case shown in Fig.\;\ref{fig:s1pics}, in the energy region 0.0-0.2 Rydbergs where there are so many resonances in the \textit{R}-matrix calculation.

The level-resolved cross sections were included from ground and metastable levels of the ionising state to all levels into the next charge state. If an upper state was not present in the atomic model, the cross section was posted to the highest state available in the ion. Only PI from ground and metastable levels were included to make the calculations more manageable. We checked the level populations for C$^0$ with and without PI included for short-lived, excited states. The fractional ion population was unchanged and it was only the very few highest levels (some of the $2s^2\,2p\,3p$ states) that were affected. This will not affect the results of the present work because the focus on is continua, and not lines.

\subsubsection{Radiative recombination}
\label{sec:rrmethods}

As previously outlined, using total PI cross sections solely to derive the rates for the reverse RR reaction in RT codes such as \lw\ presents several problems. As only a certain number of levels are included in an atomic model, the RR rates back into such limited number of levels will underestimate the total RR rate into the lower charge state as a whole. Since it is not feasible to have a large number of levels for each charge state, there has to be a `top-up' to account for the difference between the RR rates calculated from the cross sections in the model and what the total RR rate should be. Another problem is that total PI cross sections include ionisation to all levels in the next higher charge state, but are usually posted to the ground level. In this case, the reverse rates from this level can be potentially be overstated and the rates from the other levels understated, depending on the data used and electron densities. Examples of these cases are given for Si in Sect.\;\ref{sec:si_proc} and for S in Sect.\;\ref{sec:s_results}. This has the potential to distort the ion balance, which can then affect the continua and line emission calculated subsequently.

For the present work, to circumvent these problems, all recombination rates are resolved by initial and final level. The RR rates into ground and metastable levels are calculated automatically by \lw\ from the level-resolved PI cross sections. RR rates into short-lived levels have been included via the \lw\ collision rate matrix; the data sources are listed in Appendix\;\ref{sec:dr_suppress}. This potentially under-estimates contributions to the continuum from recombination into short-lived levels. However, the wavelengths of almost all these transitions into neutrals are in the visible and infra-red region. Recombination into such levels in singly-charged ions almost entirely take place below 900\,\AA. Both wavelength ranges are outside the scope of this paper; in addition, the contributions to the continuum from such transitions are expected to be negligible. Regarding the `top-up' to RR, the tendency in RT codes is to post additional rates to the ground. This could also affect level populations, including metastable levels, depending on the channels through which radiative cascades take place. Consequently, we distribute the `top up' amount to all levels included in the lower charge state in proportion to the total radiative decay rate of each level. A better alternative would be to track the channels through which decays take place from highly excited levels, but the large scale models required for this are not available.

\subsubsection{Dielectronic recombination} 
\label{sec:drmethods}

The \citet{burgess1965dr} general formula for DR, which was developed for the solar corona, can reproduce DR rates for many transition region ions. What it cannot account for is the low temperature component that occurs when EIE amongst low-lying levels is followed by capture of a free electron to levels just above the ionisation limit. Very few detailed calculations for this particular process exist, and yet it can dominate over the radiative processes PI and RR. Level-resolved DR rate coefficients were calculated for all H- to P-like ions of H to Zn, as part of the DR Project \citep[see][for the first work in the series]{badnell2003}, and have been made available on the APAP website. We use these rates in all cases and they are detailed in Appendix\;\ref{sec:dr_suppress}. As with RR, DR rates into levels higher than those included in each charge state are summed and then distributed amongst the included levels according to the total radiative decay rate of each level.

DR suppression can occur when dielectronic capture to auto-ionising states is followed by radiative stabilisation to highly-excited Rydberg states. If electron densities are high enough these states can be rapidly ionised before further radiative decays take place. This is an important process in medium-to-high density plasma, such as the solar atmosphere and fusion plasma. This is a complex process to model because of the number of states which need to be included in the modelling. Shortcuts have been developed by using the formulation of \citet{nikolic2018} or the effective recombination rates of \citet{summers1972,summers1974}. Here we suppress the DR rates based on the rates of \citet{summers1972}. The reasons for this and an illustrative case are given in Appendix\;\ref{sec:dr_suppress}.

\subsubsection{Charge transfer}
\label{sec:ct}

This process occurs during atom-ion and ion-ion collisions, where the colliders form a quasi-molecular state during the collision and an electron is shared between them. After the collision, the electron may end up in a different state, making this an excitation, ionisation or recombination process. The data most commonly used in RT codes was obtained from the Landau-Zener method. Its assumptions mean it is most appropriate for high temperature scenarios, typically 10$^5$--10$^7$\,K \citep{bates1962ct}. At low temperatures, it sometimes provides only an order of magnitude estimate. These calculations form the basis for rates included in, for example, the ion balances of \citet{arnaud1985}.

Charge transfer is another example where collisional processes can dominate over radiative processes in low temperature scenarios \citep[as discussed in][]{field1971}. Quantum mechanical, close coupling calculations using molecular orbitals provide the most accurate calculations to-date and are more appropriate for low energy collisions. The charge transfer data collated in \citet{dufresne2020pico,dufresne2021picrm} and made available in the latest version of \ch\ \citep{dufresne2024v11} have been used here; the data sources are listed in Appendix\;\ref{sec:ct_refs}.

\subsubsection{Electron collisional ionisation}
\label{sec:eii}

RT codes typically include electron impact ionisation (EII) rates calculated with approximate formul\ae. We import the electron impact ionisation rates stored in \ch\ from \citet{dere2007} for the ground levels of every charge state in the new models. For all excited levels we follow the same procedure as \citet{dufresne2021picrm}. In summary, this involves estimating the change in the excited level rates relative to the ground level rates using the \citet{burgess1983} approximation, and adjusting the \citet{dere2007} rates by this difference. 

\begin{figure*}
\centering
\includegraphics[width=1.0\textwidth]{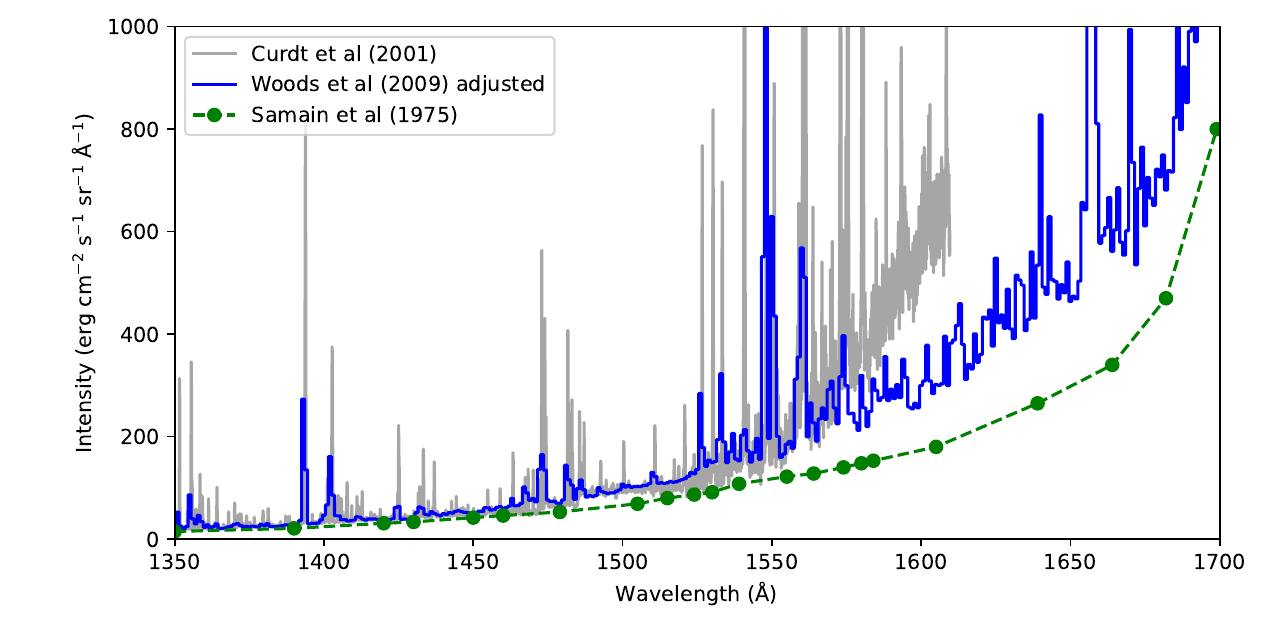}
\caption{SUMER quiet Sun disk spectrum, together with the \citet{samain_etal:1975} disk centre continuum values and the solar minimum SORCE SOLSTICE spectrum converted to disk centre, (see text for details of the conversion).}
\label{fig:obs}
\end{figure*}

\lw\ automatically calculates the reverse, three-body recombination rates from the electron impact ionisation rates. Following the discussion in Sect.\;\ref{sec:rrmethods}, the EII rates should be resolved by final level so that individual, three-body recombination rates from ground and metastable levels in the higher charge state are correctly treated, but here they are treated as total rates and posted to the final ground level. It should be noted, however, that the influence of three body recombination typically occurs through the Rydberg states very close to the ionisation limit, and these levels were not included in the models. The \ion{S}{ii} model from \citet{dufresne2021picrm}, which is discussed in Appendix\;\ref{sec:dr_suppress}, shows that three body recombination begins to influence the recombination rates at densities of 10$^{12}$\,cm$^{-3}$ and above in the chromosphere. This means the process is more likely to have an effect at the higher densities present in active regions and flares.

\subsubsection{Electron impact excitation and radiative decay}
\label{sec:levels}

Only the most advanced calculations include the resonances that appear in EIE calculations, particularly for low charge states. Examples of such theoretical methods include the \textit{R}-Matrix \citep{burke1971rm} and convergent close coupling \citep{bray1996} approaches. Resonances can make significant contributions to the EIE rate coefficients. For example, \citet{fontenla2014} used the cross sections from the \textit{R}-Matrix calculation of \citet{wang2013} for C$^0$, and the transition rate between the ground and the $2s^2\,2p^2\;^1D$ metastable level was 100 times greater than earlier values derived from simpler methods. They stated this change had a significant impact on the continuum at 1238\,\AA.

For the present work we have imported as much atomic data from the \textsc{CHIANTI} database as possible, as it includes a selection of data calculated using more advanced methods. All energy levels, radiative decay rates and EIE rates have been imported from the latest version \citep[v.11,][]{dufresne2024v11}. Many of the old atomic models include only neutrals and the ground of the singly-charged ion. While this may work for continua formed at low temperatures, it risks producing emission from higher in the atmosphere because the higher charge state will be present up to the highest temperature in the model atmosphere. Furthermore, metastable levels are required in each charge state because they influence the ion balance \citep{nussbaumer1975}. To make sure ions form at correct heights in the atmosphere we included the following ions in the calculation: \ion{C}{i-v}, \ion{Si}{i-v} and \ion{S}{i-vii}. Lines with a radiative decay rate greater than 10$^{-5}$ of the total decay rate from that upper level were included in the radiative transfer calculation for all of these ions. Thus, their effect is included in the photo-ionisation integrals. Similarly, all lines present in the older atomic models are also included in the integrals.

\subsection{Observations used for comparison}
\label{sec:obs}

There are surprisingly few observations with absolute intensities of the solar UV continuum. Aside from a few older measurements at a few wavelengths, the main instrument capable of providing accurate measurements was the Solar and Heliospheric Observatory (SOHO) Solar Ultraviolet Measurements of Emitted Radiation (SUMER, see \citealt{wilhelm1995}).

As a baseline quiet Sun, we have adopted the same reference spectrum used by many previous authors, the \citet{curdt2001} averaged, disc-centre SOHO SUMER spectrum. The main spectrum was obtained from a single slit exposure not far from Sun centre, by merging several windows in the central part of detector B, until 1485\,\AA. A different observation from detector A was added to cover longer wavelengths. As pointed out by \cite{curdt_etal:2022}, the longer wavelengths observed by detector A are partially affected by the second order hydrogen Lyman continuum, which is not simple to correct. Unfortunately, \citet{curdt2001} did not provide enough information on the data processing and the calibration used. We have re-analysed the same main observation and were unable to exactly match the published data. More details will be provided in a follow-up paper, where comparisons with other observations and a discussion on the continuum variability will be presented.

At the longer wavelengths, there are several excellent irradiance measurements, such as the SORCE Solar Stellar Irradiance Comparison Experiment (SOLSTICE), with about 1\% accuracy in the absolute radiometric calibration. A SOLSTICE spectrum during solar minimum at wavelengths above 1160\,\AA\ and resolution of 1\,\AA\ was included in the composite irradiance spectrum by \cite{woods2009}. However, to convert irradiances to radiances at disk centre is not trivial because in the UV there is both limb darkening and brightening, depending on the wavelength. 

We explored several available observations of the continua at the wavelengths of interest here and, surprisingly, we only found a single, excellent observation taken with a sounding rocket in 1973. UV radiances and the limb darkening/brightening for the continuum at different wavelengths were measured \citep{samain_etal:1975,samain:1979} using a single slit crossing the entire solar disc. The measurements were extrapolated to provide fluxes, that is, irradiances. The absolute calibration was not deemed very accurate (about 25-30\% in the region of interest here), so the relative limb darkening/brightening measurements are more accurate. We have taken these values and applied them to the SORCE SOLSTICE solar minimum spectrum to obtain an equivalent disc-centre spectrum. This spectrum is shown in Figure~\ref{fig:obs} together with the SUMER spectrum. As the extrapolation to the irradiances has in itself an additional uncertainty, we also plot the \cite{samain_etal:1975} disc-centre continuum values. They were obtained by taking the minimum value in their medium-resolution spectrum at shorter wavelengths and the maximum values longward of 1680\,\AA. Excellent agreement is found below 1485\,\AA, validating the SUMER  detector B spectrum, whilst there is clearly a problem with the calibration of the detector A spectrum. This comparison indicates that the SUMER average spectrum is a relatively good reference spectrum for the quiet Sun for the spectral region of interest here.

\subsection{The model atmosphere used for the simulations}
\label{sec:atmos}

As is customary in many radiative transfer calculations, we use a pre-existing model atmosphere from a one-dimensional, steady state calculation to provide the temperature stratification with height, as well as the total hydrogen and electron number densities. We assessed atmospheric models from several works, including \citet{fontenla1993}, \citet{avrett2008} and \citet{fontenla2014}, by comparing the Lyman continuum with quiet Sun observations. The Lyman continuum was chosen for the comparison because the various atomic rates for H are well-established, photo-ionisation can only take place to one state, and the continua from other species contribute little in this region. These factors mean it is largely possible to assess the atmosphere independently of the modelling problems described above.

\begin{figure}
\centering
\includegraphics[width=1.0\columnwidth]{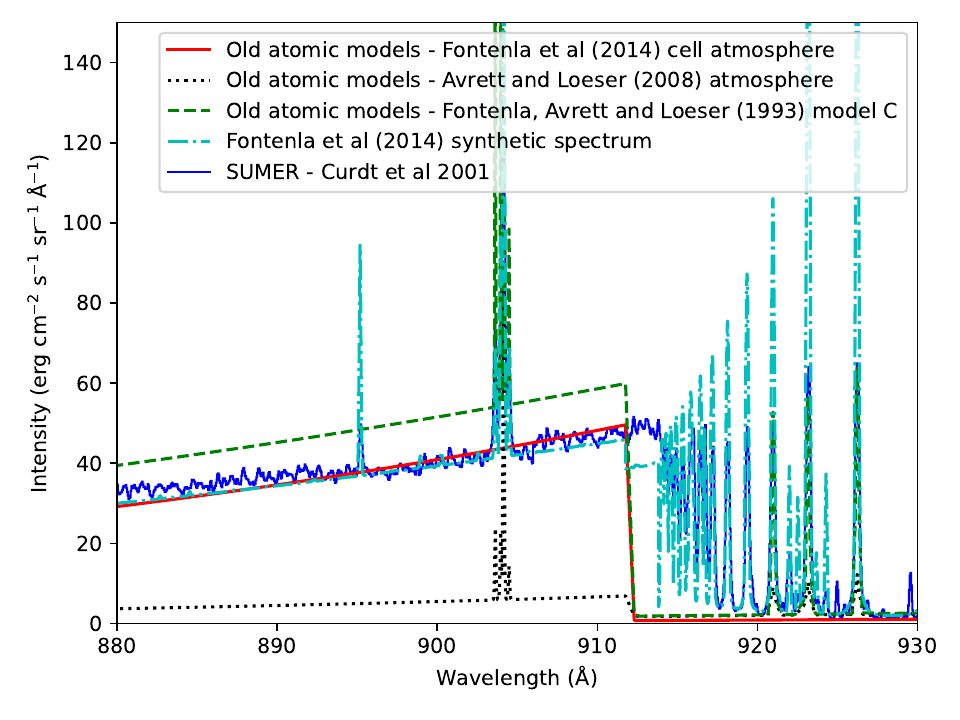}
\caption{Comparison of synthetic spectra of the Lyman continuum using various model atmospheres. The base set of atomic models was used for all of these calculations. They are compared with the synthetic spectrum for the cell model calculated by \citet{fontenla2014} and with SUMER observations.}
\label{fig:lymancont}
\end{figure}

For the test, the base set of models were run using each of the model atmospheres. From \citet{fontenla2014} we ran the cell centre and network models, (denoted in the original work by models 1301 and 1302, respectively). The \citet{fontenla2014} cell model and \citet{fontenla1993} model C reproduce the observed Lyman continuum most accurately compared to the average, quiet Sun observations used here, as shown in Fig.\;\ref{fig:lymancont}. The shape of the continua is slightly different than observations below 890\,\AA\ and again below 860\,\AA. While the \ion{N}{i} ground, \ion{Al}{ii} $3s\,3p\;^3P$ and \ion{Fe}{ii} $3d^6\,4s\;^4D$ metastable continuum edges could contribute here, the model atmospheres indicate these make virtually no contribution to emission. By contrast, the continuum derived from the \citet{fontenla2014} network model matched neither the shape nor the intensity, being more than a factor of two more intense than observations. This is perhaps not surprising because roughly 80 per cent of the solar disc, where the Lyman continuum forms, is made up of cells and the majority of the continuum emission is expected to come from such areas \citep{mariska1992}.

\lw\ reproduces very well the synthetic spectrum produced by \citet{fontenla2014} for the cell model. This is also shown in Fig.\;\ref{fig:lymancont}. However, it was not possible to tell why the intensities from \lw\ using the \citet{avrett2008} atmosphere are so low; the intensities in their work show better agreement with observations. Based on the outcome of this exercise, we use the \citet{fontenla2014} cell model throughout the rest of this work, although it is noted that a blend of this spectrum with that of the network model may give an improved picture. 

We tested computing the electron densities self-consistently during the calculation, instead of using those from the model atmosphere. However, it was only possible to do this iteration for hydrogen in the \lw\ code due to the poor conditioning of the Newton-Raphson matrix with the trace populations of some levels when all elements are considered. \citet[][their Fig. 47]{vernazza1981} show that Mg, Si and Fe make the most important contributions to electron density in the temperature minimum region, where the continua relevant for this work form. Consequently, we used the published electron densities throughout the work.

\begin{figure}
\centering
\includegraphics[width=1.0\columnwidth]{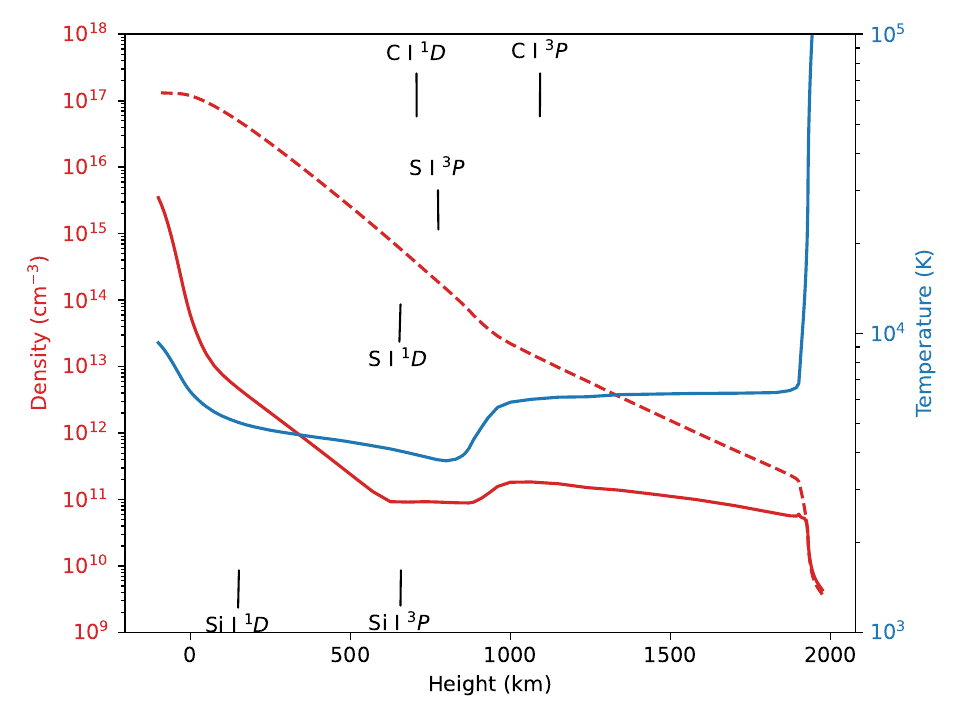}
\caption{Plot of the electron number density (red solid line), total hydrogen number density (red dashed line), and temperature (blue solid line) from the \citet{fontenla2014} cell model atmosphere. Also shown are the heights of the $\tau=1$ layer for each of the continua from the base models (black vertical lines).}
\label{fig:tempdens}
\end{figure}

\section{Results}
\label{sec:results}

The Si continua in the solar UV region mark an important transition from the primarily blackbody spectrum which forms in the photosphere in the near-UV (NUV), optical and IR, to the far- and extreme-UV and X-ray spectrum, which forms in non-LTE in the solar atmosphere. While the NUV spectrum shows obvious signs of absorption of the blackbody emission, it is the \ion{Si}{i} $3s^2\,3p^2\;^1D$ edge at 1682.12\,\AA\ that marks the point below which lines and continua are mainly in emission. This occurs on account of these features primarily being formed from the temperature minimum region and upwards. Also, the lines and continua no longer show limb darkening below this edge. At wavelengths below the \ion{Si}{i} $3s^2\,3p^2\;^3P$ edge at 1520.96\,\AA, which also forms in the temperature minimum region, lines and continua exhibit limb brightening as temperature increases with height through the atmosphere \citep{noyes1971}.

The UV continuum down to at least 1400\,\AA\ is dominated by Si, and it still influences the spectrum even below the C continuum edge at 1101\,\AA. This makes it especially important to model this element correctly in solar UV calculations. Silicon ion formation is affected by photo-ionisation, charge transfer, radiative and dielectronic recombination. It, therefore, provides a useful case to illustrate how updated modelling for each of these processes can affect the ion fractions and spectrum. Each of these effects are described in Sect.\;\ref{sec:si_proc}. Such a comparison makes it easier to understand how the processes influence the results when they are combined together, including how such processes affect the other continua. The final results with all the processes combined together are shown in Sect.\;\ref{sec:improvedresults}.

For convenience, Table~\ref{tab:edges} indicates the photo-ionization edges corresponding to the main elements and states affecting the solar UV continuum. Figure\;\ref{fig:tempdens} illustrates the variation of temperature and electron and total hydrogen number densities with height in the \citet{fontenla2014} cell model atmosphere. It also shows the height at which $\tau=1$ in the base models for each of the continua discussed here.

\subsection{Illustration of the main effects: silicon}
\label{sec:si_proc}

\begin{table*}
\caption{Ground state and main metastable terms contributing to the UV continuum.  $\lambda$  indicates the photo-ionization edge in \AA\ corresponding to the indicated level of the ground and metastable terms. The metastable terms have the same configuration as the ground unless otherwise indicated.} 
\centering	
\begin{tabular}{llrlrlrlll}
\hline\hline \noalign{\smallskip}
Element & Ground state &  $\lambda$ &  1st metastable & $\lambda$ & 2nd metastable &  $\lambda$ \\
\noalign{\smallskip}\hline\noalign{\smallskip}
C I & $2s^2 \, 2p^2$   $^3P_0$ & 1101.07  &  $^1D_2$ & 1240.27 &  $^1S_0$ & 1445.66   \\ 

N I & $2s^2 \, 2p^3$  $^4S_{3/2}$ & 853.06 & $^2D_{5/2}$ & 1020.40 & $^2P_{1/2}$ & 1131.39  \\ 

O I & $2s^2 \, 2p^4$  $^3P_2$ & 910.44 & $^1D_2$ &	1064.18	&  $^1S_0$ & 1315.02  \\ 


Na I & $3s$  $^2S_{1/2}$  & 2412.58 & &  \\  

Mg I & $3s^2$   $^1S_{0}$ & 1621.51 & $3s \,3p$ $^3P_0$ & 2511.26 &  \\  

Al I &  $3s^2 \, 3p$  $^2P_{1/2}$ & 2071.32  & $3s \,3p^2$ $^4P_{1/2}$ & 4360.98 & \\ 

Si I & $3s^2 3p^2$ $^3P_0$ & 1520.96    & $^1D_2$ & 1682.12 & $^1S_0$ & 1985.96  \\  

S I & $3s^2 \, 3p^4$  $^3P_2$ & 1196.76 & $^1D_2$ & 1345.52 & $^1S_0$ & 1629.22  \\ 

 K I & $3p^6\, 4s$   $^2S_{1/2}$ & 2856.34 &   &  \\ 

 Ca I & $4s^2$ $^1S_{0}$ & 2028.15 & $4s\, 4p$ $^3P_0$	& 2928.42 & $3d \,4s$  $^3D_1$ & 3451.78 \\  
 
 Mn I & $3d^5 \, 4s^2$   $^6S_{5/2}$  &  1667.80 & &  \\  

 Fe I &  $3d^6 \, 4s^2$   $^5D_{4}$  & 1568.95  & $3d^7 \, 4s$ $^5F_5$  & 1760.29   \\ 

 Ni I & $3d^8 \, 4s^2$   $^3F_{4}$ & 1623.38 & $3d^9 \,4s$ $^3D_1$ & 1669.81 & $3d^9 \,4s$ $^1D_2$  & 1718.51  \\ 

\hline\hline \noalign{\smallskip}
\end{tabular}
\label{tab:edges}
\end{table*}

\subsubsection{Collisional processes}

\begin{figure}
\centering
\includegraphics[width=\columnwidth]{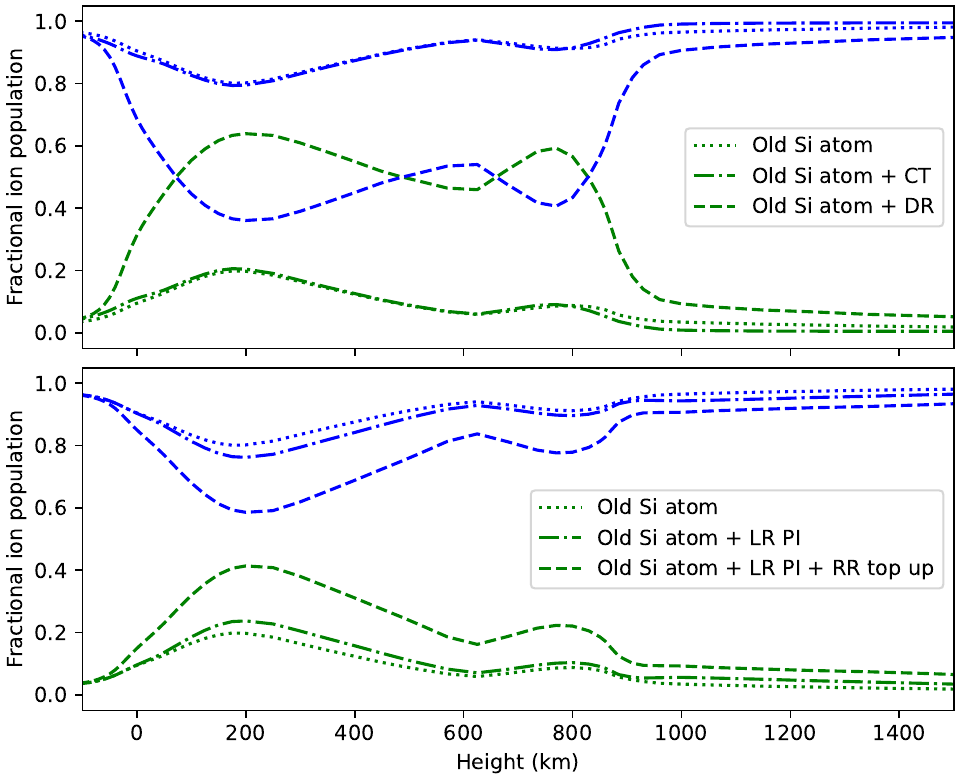}
\caption{Comparison of Si ion fractions derived from the base set of models when different atomic processes are each added individually to the old Si model. Green lines denote Si$^0$ and blue lines denote Si$^+$. The top plot shows the collisional processes and the bottom plot shows the radiative processes.}
\label{fig:si_base_ionfracs}
\end{figure}

\begin{table}
\caption{Comparison of total ionisation rates (in s$^{-1}$) for each process from all terms in the 3s$^2$\,3p$^2$ ground configuration in Si$^0$ at the temperature minimum (3750\,K). The fractional population of each term at the temperature minimum is also given.} 
\centering	
\begin{tabular}{p{1.3in}ccc}
\hline\hline \noalign{\smallskip}
Process & $^3$P & $^1$D & $^1$S \\
\noalign{\smallskip}\hline\noalign{\smallskip}

Old atomic models  \\
PI & 0.538 & 3.25 & 302 \\
CT & - & - & - \\
CI & 10$^{-8}$ & 10$^{-7}$ & 10$^{-5}$ \\
\noalign{\smallskip}
New atomic models \\
PI & 0.537 & 3.55 & 466 \\
CT & 1.81 & 3.35 & - \\
CI & 10$^{-8}$ & 10$^{-7}$ & 10$^{-5}$ \\
\noalign{\smallskip}\hline
Fractional population & 0.9497 & 0.050 & 0.0003 \\
\hline\hline \noalign{\smallskip}
\end{tabular}
\label{tab:ionrates}
\end{table}

\begin{table}
\caption{Comparison of total recombination rates (in s$^{-1}$) for each process from the 3s$^2$\,3p\;$^2$P ground term and 3s\,3p$^2$\;$^4$P metastable term in Si$^+$ at the temperature minimum (3750\,K). The fractional population of each term at the temperature minimum is also given.} 
\centering	
\begin{tabular}{p{1.3in}cc}
\hline\hline \noalign{\smallskip}
Process & $^2$P & $^4$P \\
\noalign{\smallskip}\hline\noalign{\smallskip}

Old atomic models  \\
RR & 0.084 & - \\
DR & - & - \\
CT & - & - \\
Three body & 10$^{-5}$ & - \\
\noalign{\smallskip}
New atomic models \\
RR from PI & 0.083 & 0.056 \\
RR top-up & 0.027 & 0.004 \\
DR & 0.115 & 0.175 \\
CT & 0.00 & 63000 \\
Three body & 10$^{-5}$ & - \\
\noalign{\smallskip}\hline
Fractional population & 1.00 & 10$^{-6}$ \\
\hline\hline \noalign{\smallskip}
\end{tabular}
\label{tab:recrates}
\end{table}

To start with, we consider the treatment of charge transfer and dielectronic recombination in Si ion formation. We add these processes one at a time to the old \texttt{Si\_atom} to understand their effect. The new CT and DR rates added to the model are resolved by initial and final levels. In each case, the relative populations of the ground and metastable levels do not change, indicating that collisional (de-)excitation rates from the metastable levels dominate at the higher densities of the temperature minimum region. Since the processes considered in this section are collisional processes the changes that occur in the ion balance and continua are entirely caused by changes to the ionisation and recombination rates added to the Si model.

\begin{figure*}
\includegraphics[width=\textwidth]{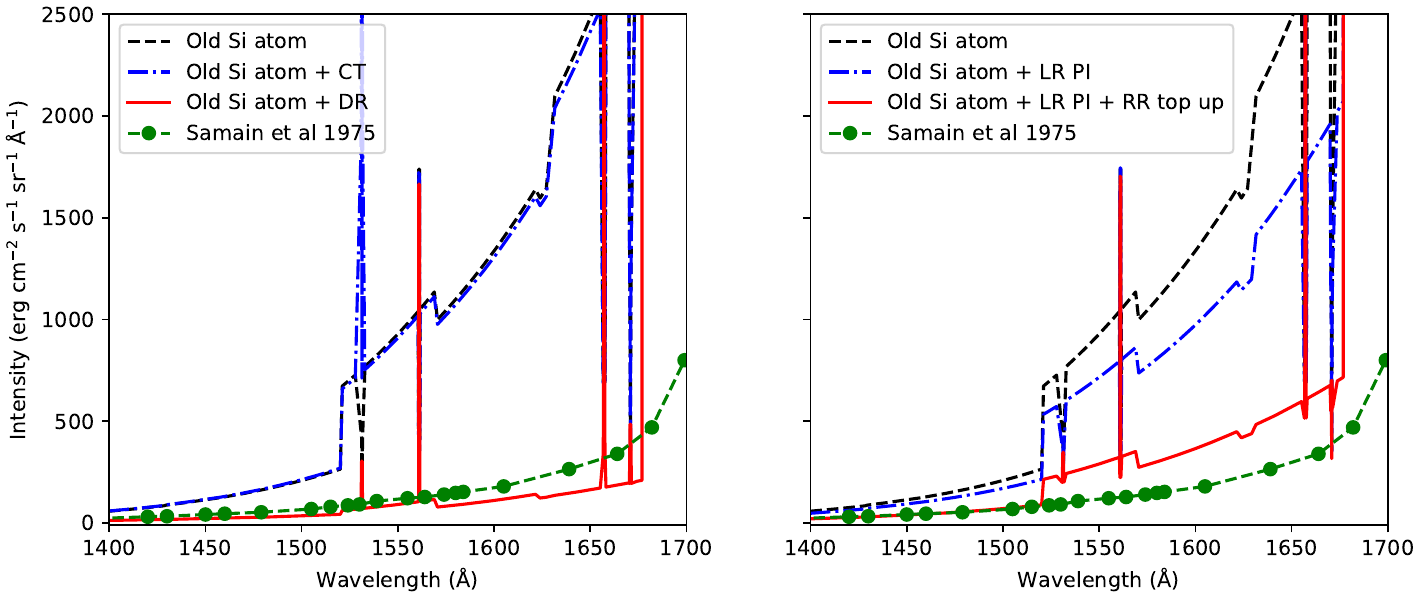}
\caption{Comparison of Si continua derived from the base set of models when different atomic processes are each added individually to the old Si atomic model. The left plot shows the collisional processes while the right plot shows the radiative processes.}
\label{fig:si_base_cont}
\end{figure*}

Figure\;\ref{fig:si_base_ionfracs} shows the fractional populations of Si$^0$ and Si$^+$ from the base set of models compared with the populations when CT and DR are added independently. (The height at which the slow temperature decrease from the photosphere begins is 200\,km and this continues to the temperature minimum region at 800\,km.) The small change in populations when CT is added belies how substantial the CT rates are. \citet{dufresne2021picrm} show that the total CT ionisation to total PI rates from the ground level at 10,000\,K are in the ratio 28:1. In the present work, the same ratio is 37:1, which is in relatively good agreement considering the difference in model atmospheres used and the radiative transfer treatment here. The CT ionisation rate coefficients drop significantly at the temperature minimum region, plus the protons required for this process are more scarce.  However, the CT ionisation rates are still very important at these temperatures, as illustrated in Table\;\ref{tab:ionrates}. Taking into account the fractional population of each term, the overall ionisation rate is almost a factor of four higher when this process is included.

The only important CT recombination reaction is from the metastable term in Si$^+$. Since the process involves neutral H it explains the substantial rate shown in Tab.\;\ref{tab:recrates}. The process contributes an extra 67\% to the overall recombination rate from this ion despite the population of this term being so low at these temperatures. (There is no RR in the old atomic model from the $3s^2\,3p\;^4$P metastable term, for reasons which will be discussed in the next section.) As a whole, there is almost no change to the ion fractions and in the spectrum, as shown in Fig.\;\ref{fig:si_base_cont}.

The old Si model has no DR rates in it at all, including from Si$^{+2}$ to Si$^+$. The APAP DR rates from Si$^+$ have a substantial component of DR at low temperatures, and it is comparable to the RR rates at these temperatures. This explains the significant depopulation of Si$^+$ when DR is included. The effect this has on the solar spectrum is clear to see in Fig.\;\ref{fig:si_base_cont}. Because this is a collisional process, the greater presence of neutral Si means there is substantial absorption in the $^1$D continuum, the edge of which is at 1682.1\,\AA. The depletion of this continuum means that the edge is now barely visible in the $^3$P threshold at around 1520\,\AA. The \ion{S}{i} absorption edge at 1629\,\AA\ disappears now and the \ion{Fe}{i} edge at 1569\,\AA\ and the \ion{Mg}{i} edge at 1621\,\AA\ are shallower. Compared to this, the continua from the base models are so intense that lines down to 1450\,\AA\ are in absorption, which clearly contradicts observations.

\subsubsection{Radiative processes}

Figure\;\ref{fig:si_base_ionfracs} shows the effects on the Si ion balance when new radiative processes are added to the base set of models. As known from \citet{lanzafame1994} and \cite{avrett2008}, Si$^0$ is highly depleted through the whole atmosphere as a result of photo-ionisation. When level-resolved photo-ionisation cross sections are used instead of the OP total cross sections the difference in the ion fractions is small. This is primarily because the total photo-ionisation rates out of the ion for the ground and first metastable levels are very similar, as indicated by the total cross sections in Fig.\;\ref{fig:si1pics}. When tested on a blackbody spectrum the PI rates for the level resolved cross sections and those available in \lw\ differ by only 6\% for the ground and 13\% for the first metastable. There is a larger difference for the second metastable term, though, which explains the difference in rates in Tab.\;\ref{tab:ionrates}. PI from the ground and metastable levels of Si$^0$ is dominated by ionisation to the ground of Si$^+$. The next highest states in Si$^+$ to which PI takes place is at wavelengths below 1000\,\AA\ and the continuum is much weaker there than in the 1500-1600\,\AA\ region. Consequently, here, the differences in the ion fractions and continua between using level resolved and total cross sections are caused by differences in the cross section values of the OP and AS calculations, and not from the way PI is treated.

There is one example in the old Si model of how PI cross sections can be handled incorrectly. In the old atomic model, PI rates from all the Si$^0$ states are posted to the Si$^+$ $^2P$ ground. PI from the $3s\,3p^3\;^5S$ state only takes place to the Si$^+$ $3s\,3p^2\;^4P$ metastable state. The cross section included is hydrogenic and because it is posted to the Si$^+$ ground the transition has a threshold of 3066.7\,\AA\ in the old model. In reality, it should have a threshold of 1304.2\,\AA. This will affect both the PI rate out of the level and the wavelength at which RR contributes to the spectrum. In this particular example, however, the mistake does not impact on the results from the base models.

The main shortcoming in the treatment of RR is that it relies solely on using the PI cross sections included in the model. In any model there will inevitably be a limit to how many levels in the lower charge state can be included. The result is that RR into more highly-excited states will be missed and this can make a large contribution to the total recombination rate. It is seen in Tab.\;\ref{tab:recrates} that this missing contribution accounts for a quarter of the total rate from the ground. This makes a striking difference to the ion fractions shown in Fig.\;\ref{fig:si_base_ionfracs}.

\begin{figure}
\includegraphics[width=\columnwidth]{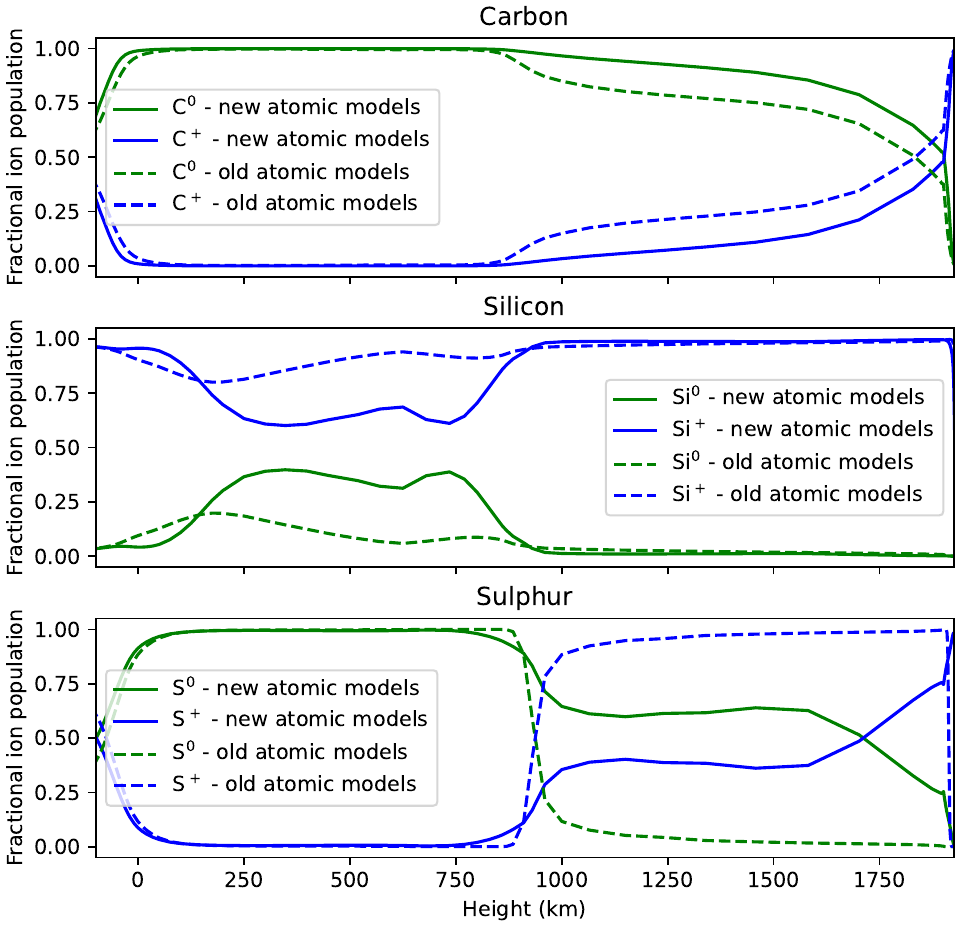}
\caption{Comparison of the ion fractions for C, Si and S derived from the base set and new set of atomic models.}
\label{fig:ionfracs_all}
\end{figure}

\begin{figure*}
\centering
\includegraphics[width=\textwidth]{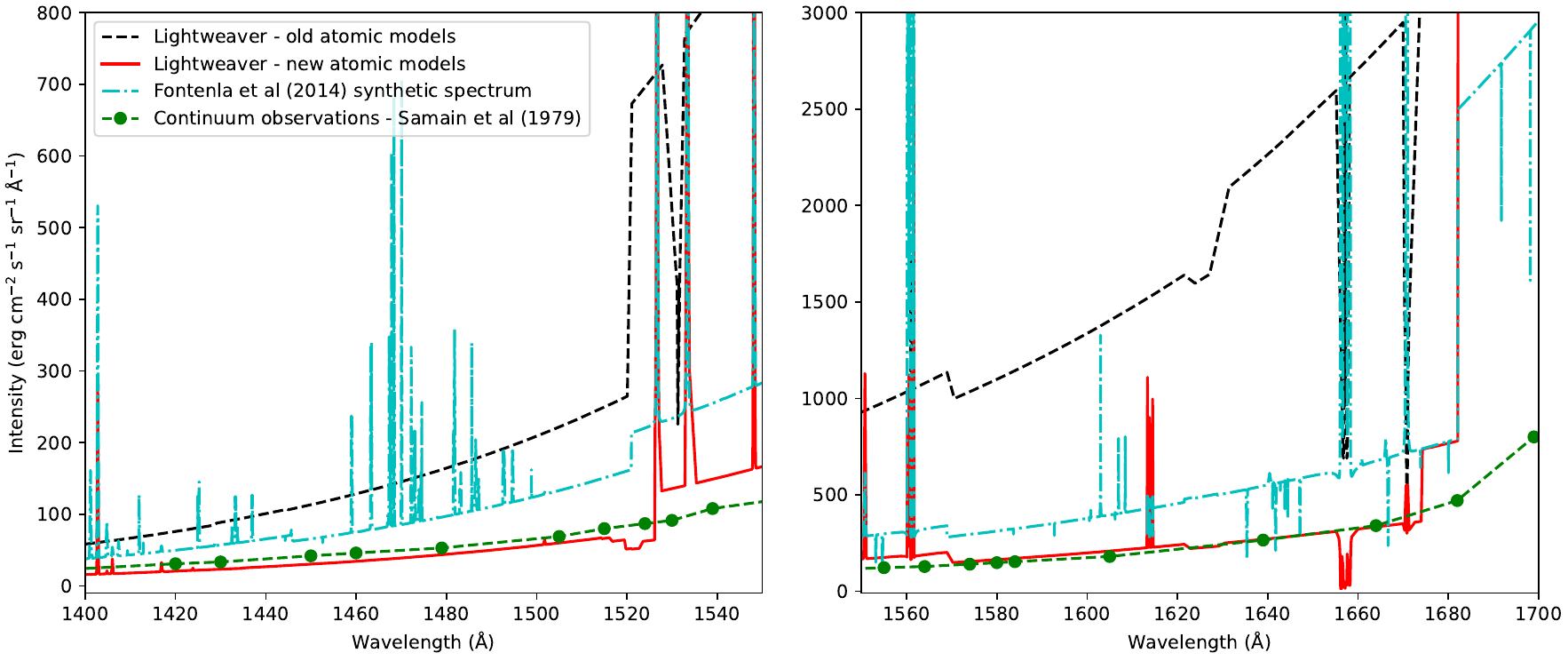}
\caption{Comparison of Si continua from various synthetic spectra with observations.}
\label{fig:si_cont}
\end{figure*}

Based on the above discussion, Fig.\;\ref{fig:si_base_cont} shows that the addition of only level resolved PI produces a relatively small change in the continuum at the $^3P$ edge. The small difference in the ion fractions at lower heights, where the $^1D$ continuum forms, produces a slightly more pronounced change in the continuum at wavelengths shorter than 1682.1\,\AA. Topping up the RR rates, and the extra absorption of the continuum caused by the higher population of the Si$^0$ $^1D$ term, brings the synthetic spectrum into much better agreement with observations. As seen in the previous section, it still takes DR to flatten the $^3P$ edge, as well as the \ion{Fe}{i} 1569\,\AA, \ion{Al}{i} 1621\,\AA\ and \ion{S}{i} 1629\,\AA\ edges, to make it more consistent with observations.

\subsection{The new set of atomic models}
\label{sec:improvedresults}

Now that the effect of atomic processes on the synthetic spectrum have been highlighted individually, we move on to showing the final results obtained when the new models are used for the whole calculation. The changes made to the models are described in Sect.\;\ref{sec:atoms} and they are compared to the results obtained when the base set of models is used for the calculation. All other parameters, such as electron and total hydrogen densities, micro-turbulent velocities, temperature, are kept the same so that the changes are created exclusively by the new atomic models. Since we use the model atmosphere from \citet{fontenla2014}, we also compare the results with the synthetic spectrum for the cell model produced by their work.

\subsubsection{Silicon}

We have discussed in detail how the various atomic data individually affect Si, and so we omit a detailed discussion for the new model. The Si ion fractions from the base and new sets of models are shown in Fig.\;\ref{fig:ionfracs_all}. Comparison with the ion fractions in Fig.\;\ref{fig:si_base_ionfracs} shows that, as expected, DR and the top up to RR are the greatest influence on the ion balance. However, when all processes are combined together Si$^0$ is not as populated as when DR is included alone, which indicates that CT ionisation does offset the new recombination processes to a certain degree.

The \ion{Si}{i} continua produced by the new models are shown in Fig.\;\ref{fig:si_cont}. Although there is some uncertainty in the absolute intensities of the \citet{samain_etal:1975} observations, as described in Sect.\;\ref{sec:obs}, it can be seen how closely the new models reproduce the overall shape of both the $^3P$ and $^1D$ continua in the wavelength region shown. The $^3P$ continua edges are at 1520.96\,\AA, 1522.75\,\AA\ and 1526.14\,\AA\ and the new models show small transitions between emission and absorption at these edges. The step in the continuum at 1570\,\AA\ down to 1526\,\AA\ affects the longest Si$^0$ edge at 1526.14\,\AA. This is caused by \ion{Fe}{i} and may well flatten if treated correctly with a new atomic model. As seen in the previous section, the old atomic model has a very deep absorption edge at 1520\,\AA. This is largely because the lower population of Si$^0$ in the base models, shown in Fig.\;\ref{fig:ionfracs_all}, means that there is insufficient absorption of the blackbody radiation by the $^1D$ state at wavelengths between its edge at 1682\,\AA\ and the $^3P$ edge at 1520\,\AA. 

The results for the $^1D$ continuum show that the new models and the \citet{fontenla2014} spectrum are closer to observations. The \citet{fontenla2014} spectrum is reasonably close in absolute intensity to the disc-centre radiances we obtained from SOLSTICE, which is shown in Fig.\;\ref{fig:obs}. In the description of the radiative transfer codes \citep{fontenla2009} used to produce the \citet{fontenla2014} spectrum, an ``extra'' source of opacity and emission at the temperature minimum region was included to account for the discrepancy commonly found between radiative transfer codes and observations of the 1300-3600\,\AA\ range in the Sun and other cool stars. The physical nature of this so-called ``missing" opacity is often attributed to either a forest of absorption lines or molecular continua, both of which are often treated in LTE. \cite{fontenla2009} show that this decreases their synthetic continuum by about a factor of two in the range 1300-1600\,\AA. This factor may account for the \citet{fontenla2014} spectrum being in good agreement with observations at these wavelengths. In other words, it appears that their synthetic spectrum without ``missing'' opacity would be similar to the spectrum produced by \lw\ using the old atomic models in the range 1300-1600\,\AA\ because they use similar atomic models and \lw\ does not include any ``missing'' opacity. In the present work, we find that improving the treatment of the atomic processes alone is able to account for the observed continua all the way up to 1680\,\AA.

\subsubsection{Carbon}

The carbon continua form at heights between the silicon UV continua just shown and the hydrogen Lyman continuum. With the base models the $\tau$=1 level for the ground term (2s$^2$\,2p$^2\;^3P$) continuum is at 1150\,km (6120\,K), while the 2s$^2$\,2p$^2\;^1D$ continuum $\tau$=1 layer is at 770\,km (3785\,K). 

\begin{figure}
\centering
\begin{subfigure}{0.45\textwidth}
\includegraphics[width=\textwidth]{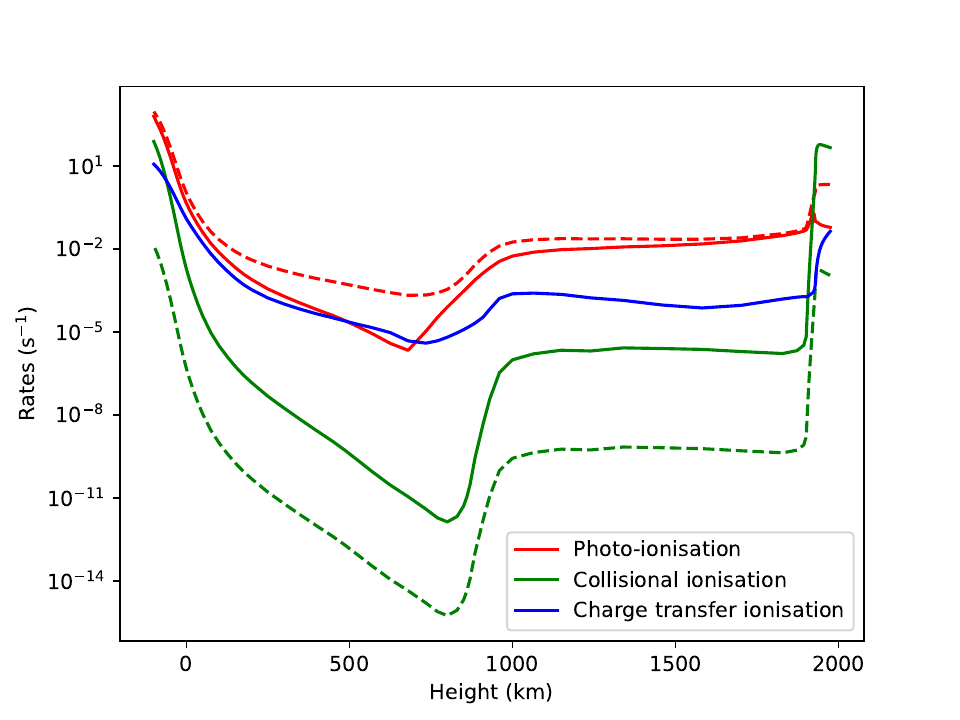}
\caption{Comparison of overall ionisation rates from the base set (dashed lines) and new set (solid lines) of atomic models. Different colours represent the different atomic process, as given in the legend.}
\label{fig:c_ionrates}
\end{subfigure}
\hfill
\centering
\begin{subfigure}{0.45\textwidth}
\includegraphics[width=\textwidth]{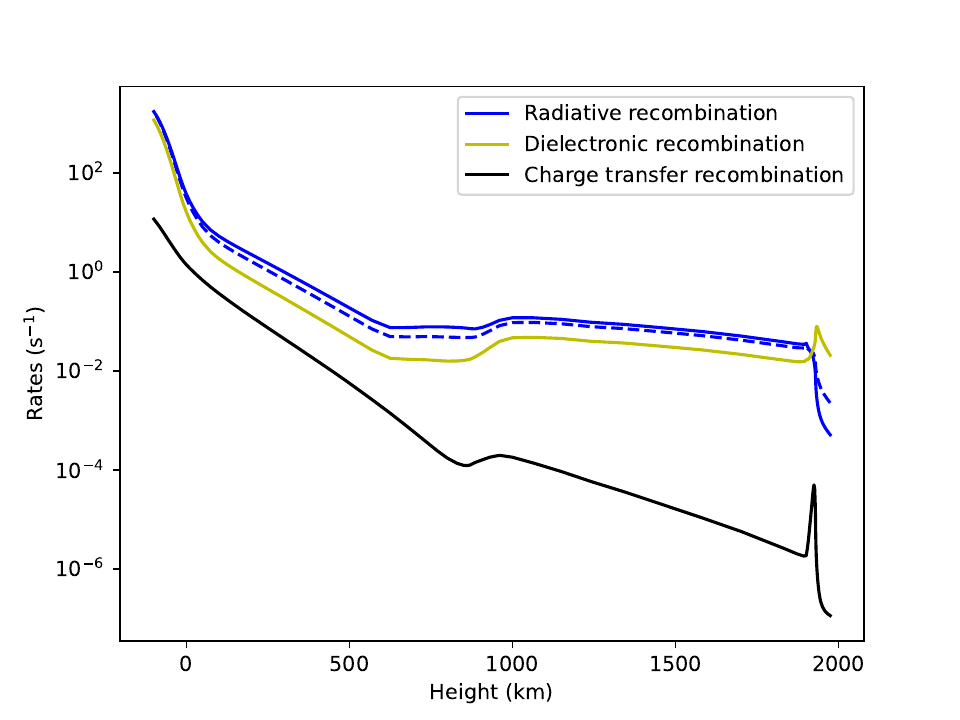}
\caption{Comparison of overall recombination rates from the base set (dashed lines) and new set (solid lines) of atomic models. Different colours represent the different atomic processes, as given in the legend.}
\label{fig:c_recrates}
\end{subfigure}

\caption{Comparison of overall ionisation and recombination rates for carbon.}
\label{fig:c_rates}
\end{figure}

The relative importance of the various atomic processes for C over the whole chromosphere is shown in Fig.\;\ref{fig:c_rates}. \citet{dufresne2020pico} found that CT was unimportant for carbon, but because they used optically thin models, constant pressure and an approximation for PI they could only test this at around 10\,000\,K. While CT does not dominate here it still makes a contribution between the photosphere and temperature minimum region in the 5-20\% range relative to PI in the older atomic model, as seen in Fig.\;\ref{fig:c_ionrates}.

As expected, the collisional ionisation rate for carbon is considerably weaker than the other ionisation processes over much of the chromosphere. However, it should be noted that the CI rates in the new models are three orders of magnitude higher than in the old atomic model. Even though \lw\ implements the \citet{burgess1983} approximation, which was developed specifically for low charge ions, it is seen that the \citet{dufresne2021picrm} approach is better for modelling because it combines the \citet{burgess1983} approximation with the \citet{dere2007} cross sections. (The latter were thoroughly checked with experiment.) CI becomes relevant at temperatures above 9000\,K (1900\,km in the model atmosphere), where the C$^0$ ion fractions start to go below 50\%, as seen in Fig.\;\ref{fig:ionfracs_all}. In the new models CI is the dominant process above that height, but in the base models PI is the sole ionising process throughout the atmosphere.

\begin{figure*}
\centering
\includegraphics[width=\textwidth]{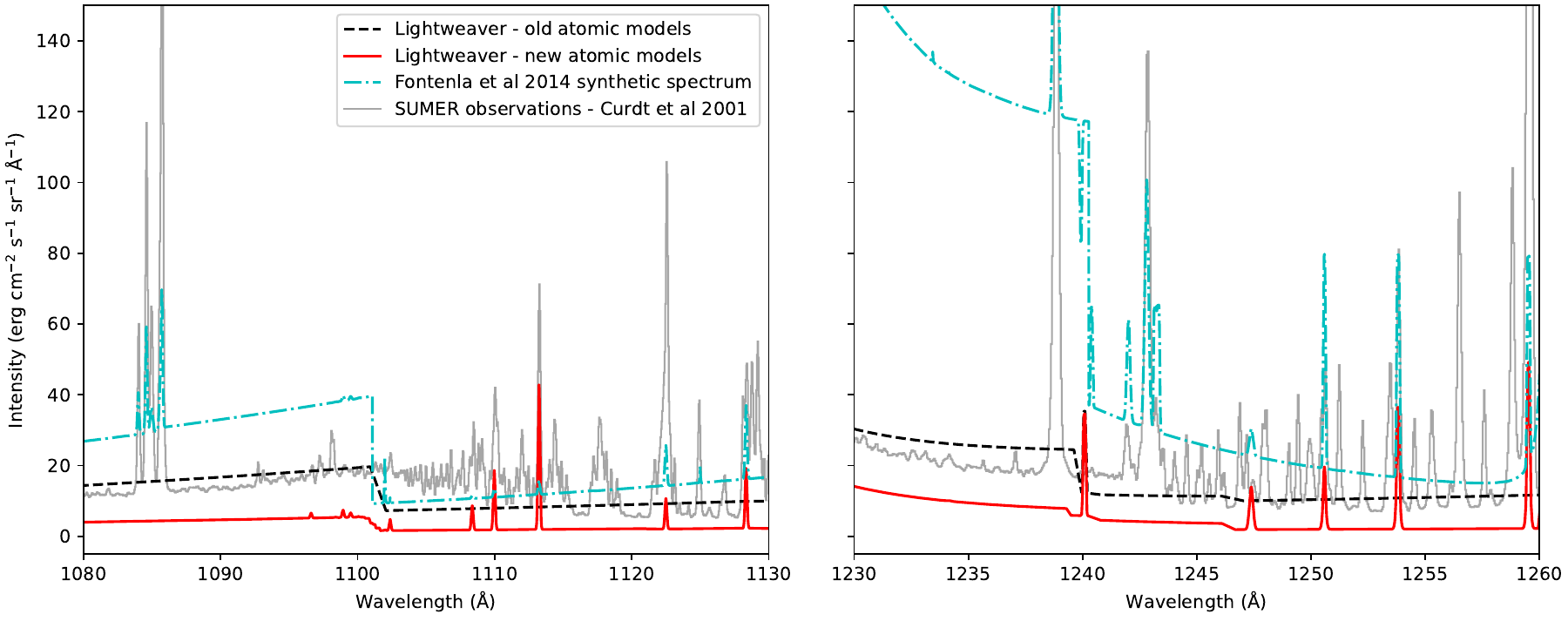}
\caption{Comparison of C continua from various synthetic spectra with observations.}
\label{fig:c_cont}
\end{figure*}

The recombination rates in Fig.\;\ref{fig:c_recrates} make it clear that CT recombination is an unimportant process for C in the solar atmosphere, even where the neutral H required for this process is abundant. DR is generally a factor of two to four lower than RR, but it is only 50\% lower than RR in the photosphere and above 9000\,K in the chromosphere. Overall then, it is still an important process to include for carbon in the lower solar atmosphere.

The final ion fractions for both sets of models, shown in Fig.\;\ref{fig:ionfracs_all}, inform the comparison of the synthetic spectrum of the C continua, which is shown in Fig.\;\ref{fig:c_cont}. Remembering that the $\tau$=1 layer for the ground term continuum with an edge at 1101\,\AA\ occurs at 1150\,km, this is where the largest difference in the ion fractions occurs. It is partly caused by the higher recombination rates described above. The DR rates from C$^+$ in \citet{altun2004,altun2004err} at chromospheric temperatures are higher than those from \citet{nussbaumer1983}. This might be a factor causing the continuum to be lower and is discussed further in Sect.\;\ref{sec:discussion}. In addition, the PI rates in the new models are lower by a factor of two compared to the base models at the $\tau=1$ layer because the Si continuum is lower by this amount in the new models. As a whole, this means that C$^+$ is less populated in this region, and the edge is shallower than in the base models and in the observations. The higher Si continuum seen in the previous section when using the base models is responsible for the continuum being higher than the SUMER observations at wavelengths above and below the edge at 1100\,\AA\ with those models. The SUMER spectrum appears to have a small head in the continuum in the region 1090-1100\,\AA, but the \citet{woods2009} continuum does not. It cannot be caused by the Rydberg transitions to the C$^+$ $^2P_{3/2}$ ionisation limit, since they end at 1100\,\AA. The \citet{parenti2005} SUMER atlas indicates it could be a small cluster of lines, but these would not be resolved in the \citet{woods2009} composite spectrum, which originates from a low-resolution spectrum at these wavelengths.

Also shown in Fig.\;\ref{fig:c_cont} is the \citet{fontenla2014} synthetic spectrum. Their \ion{C}{i} $^3P$ continuum is much higher even though they use PI cross sections with the same threshold value as the old atomic model. This means their RR rates should be the same, plus the same model atmosphere is being used. Their continuum at wavelengths longer than the edge also has a similar intensity. Therefore, the reasons for the differences in their spectrum is not obvious without a thorough comparison of the two codes, which is beyond the scope of this article. The continuum edges in all three synthetic spectra have different shapes. In the new models there is extra detail because the cross sections are resolved by fine structure, for which there are six transitions from C$^0\;^3P_{0,1,2}$ to the C$^+\;^2P_{1/2,3/2}$ ground states. In reality, these edges will never be resolved observationally because the C$^0$ Rydberg lines rising up to the edge will always obscure them.

\citet{lin2017} discuss how PI of the C$^0\;2s^2\,2p^2\;^1D$ metastable level, with a threshold at 1240.27\,\AA, caused by Lyman-$\alpha$ is an important process in the formation of some \ion{C}{i} lines. Line intensities can be enhanced when PI is followed by recombination into excited states. Looking at the continua from the present work shown in Fig.\;\ref{fig:c_cont}, the results for the \ion{C}{i} $^1D$ continuum are similar to the results highlighted for the $^3P$ continuum above. The continuum in the new models is a little too low at wavelengths above and below the edge, while it is a little too high the old models. Again, the \citet{fontenla2014} spectrum has a very steep and large drop at the edge. In that work, they question whether, amongst other things, the large collision strength from \citet{wang2013} populating the $^1D$ term from the ground could be too high. This could cause enhanced ionisation of C$^0$ and a steep $^1D$ edge following recombination. We use the same collision data here and do not have this problem. Their C abundance is within 10\% of the \citet{asplund2009} abundance used by \lw. An alternative possibility is that their broader and more intense Ly-$\alpha$ wings may be over-ionising C$^0$, which could enhance both of the $^3P$ and $^1D$ edges through recombination.

\subsubsection{Sulphur}
\label{sec:s_results}

Selection rules for transitions between S$^0$ and S$^+$ mean that there are significant rates between ground and metastable initial and final levels to illustrate the difference in using total and level-resolved PI cross sections. The S$^0$ total PI cross section from the ground, $3s^2\,3p^4\;^3P$ state in Fig.\;\ref{fig:s1pics} shows that the threshold values in the cross sections to the final metastable $3s^2\,3p^3\;^2D,\,^2P$ states in S$^+$ (with thresholds at about 0.15 and 0.20 Rydbergs, respectively) are similar to the threshold value for the transition to the final, $3s^2\,3p^3\;^4S$ ground state. When the rates for the reverse, RR reaction are calculated from the total PI cross section, these thresholds for recombination from the metastable states equate to a thermal temperature of about 24\,000\,K and 32\,000\,K, respectively. RR is a spontaneous process and so the threshold from each state should be at zero energy. The only situation in which using the total cross sections and posting to the ground level will be correct is if the two following conditions are true. The first is that the total cross sections must have their energy threshold starting at the ground level energy. The second is that the populations of the metastable levels must be determined solely by electron impact excitation and de-excitation, in other words, electron densities should be above the critical density for each level.

\begin{table}
\caption{Comparison of RR rates from the base and new sets of models (in s$^{-1}$) from all terms in the $3s^2\,3p^3$ ground configuration of S$^+$ at two different temperatures. The fractional populations of each term according to the new atomic model are also given.} 
\centering	
\begin{tabular}{p{1.3in}ccc}
\hline\hline \noalign{\smallskip}
Model & $^4S$ & $^2D$ & $^2P$ \\
\noalign{\smallskip}\hline\noalign{\smallskip}

At 3750 K  \\
\noalign{\smallskip}
Old atom & 0.058 & - & - \\
New atom & 0.037 & 0.035 & 0.045 \\
\noalign{\smallskip}
Fractional population & 0.99 & 0.01 & 10$^{-4}$ \\
\noalign{\smallskip}
\noalign{\smallskip}
At 20\,000\,K \\
\noalign{\smallskip}
Old atom & 0.022 & - & - \\
New atom & 0.009 & 0.007 & 0.009 \\
\noalign{\smallskip}
Fractional population & 0.47 & 0.41 & 0.12 \\
\hline\hline \noalign{\smallskip}
\end{tabular}
\label{tab:s_pirates}
\end{table}

To illustrate this, we present in Tab.\;\ref{tab:s_pirates} the recombination rates from S$^+$ from the ground and metastable states at two different points in the atmosphere. The first is at the temperature minimum close to where its continua form (3750\,K, 800\,km), and the second at 20\,000\,K (1925\,km) where S$^+$ is 50\% populated. Firstly, at 3750\,K the new atomic model rate from the ground is about 60\% lower than the rate from the old model. This equates to the difference between the cross sections for the two models at zero threshold, as seen in Fig.\;\ref{fig:s1pics}. It means the temperature is too low for recombination from the metastable states in the total cross section to make a contribution to the ground level rate in the older model. There are significant RR rates from the two metastable levels in the new models, but the metastable populations are too small to affect the recombination rate at the temperature minimum.

At a temperature of 20\,000\,K the metastables have a significant population. At this temperature, the old rate from the ground level is more than twice as large as the same rate in the new models because it now includes contributions from the thresholds at 0.15 and 0.20 Rydbergs. If the individual rates from each level are weighted by their populations, it can be seen that the total RR rate from the old model (0.010\,s$^{-1}$) is virtually the same as that in the new model (0.008\,s$^{-1}$). This occurs, as indicated above, because densities are high enough that the Boltzmann factor in the EIE rate populates the metastable levels in the new model in the same way that the Boltzmann factor in the RR rate includes contributions from the metastable levels in the total PI cross section from the old model. A frequently encountered situation in which using total PI cross sections is likely to break down is in the case of Mg$^0$ ionising to Mg$^+$. The cross section from the $3s^2\;^1S$ Mg$^0$ ground state to the first excited term in Mg$^+$, $3p\;^2P$, is about an order of magnitude higher than PI to the ground, $3s\;^2S$ state, according to the \citet{badnell2006} cross sections. However, the first excited term is short-lived. Therefore, if total cross sections were used, a contribution to the RR rate from the excited state in Mg$^+$ would be included in the model and overstate the recombination, whereas this state would not contribute to recombination in reality.

\begin{figure*}
\centering
\includegraphics[width=\textwidth]{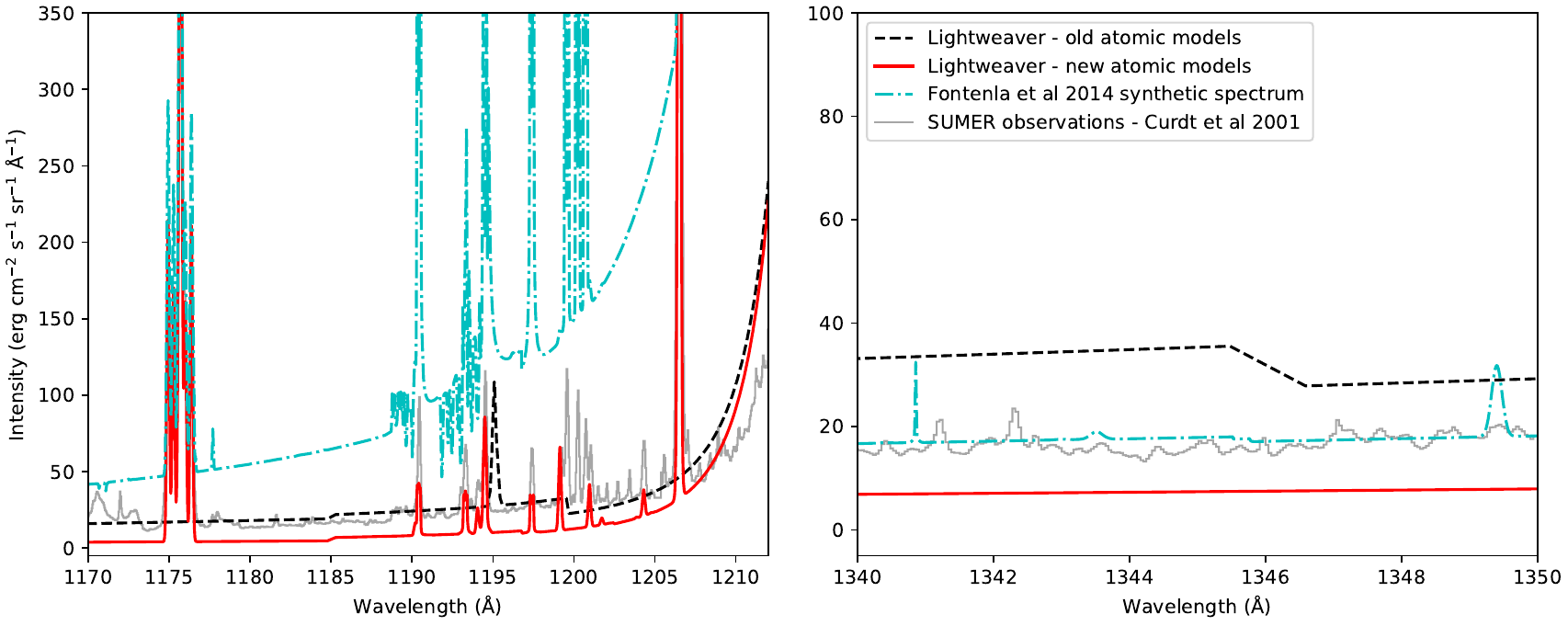}
\caption{Comparison of S continua from various synthetic spectra with observations.}
\label{fig:s_cont}
\end{figure*}

Moving on to the output for S from the models, the main difference between the two sets of models is that it can be treated in non-LTE in the new models. In the old model the atomic data is treated in $LS$-coupling. This compresses the three continuum edges for the $3s^2\,3p^4\;^3P$ ground term, which experimentally are at 1196.8,\AA, 1202.5\,\AA\ and 1205.0\,\AA, into one edge at 1199.6\,\AA. These factors, and the lower proportion of S$^+$ present because of the significant DR rates at low temperature, cause the edges in the new models to be very small by comparison. This can be seen in Fig.\;\ref{fig:s_cont}. It is difficult to compare the size of these edges with observations because: they are so small, many strong lines are present, and the difficulty in modelling the Ly-$\alpha$ wings in a hydrostatic, one-dimensional atmosphere. It is still possible to see, however, how intense and broad the Ly-$\alpha$ wings are in the \citet{fontenla2014} spectrum by comparison.

Again, the continuum at wavelengths longer than the \ion{S}{i} $^1D$ edge at 1345.5\,\AA\ in the base models is higher, as seen in Fig.\;\ref{fig:s_cont}, because of the influence of the Si continua. \citet{fontenla2014} use a PI cross section that is lower at threshold than the OP value for the $3s^2\,3p^4\;^1D$ metastable term in S$^0$ to resolve an excessively high continuum. While old atomic model uses the OP cross section, it produces an higher edge primarily because it is treated in LTE. The cross section in the present, level-resolved model are similar to the OP data, but the continuum has a very shallow edge. This shows that treating this and other atomic processes correctly can produce an edge which is in better agreement with observations, without needing to adjust existing atomic data.

\section{Discussions and future improvements}
\label{sec:discussion}

The present work clearly shows what improvements can be obtained with solar UV radiative transfer calculations when updated atomic models are included. There are a number of issues that we have encountered during the present work and which have implications for existing and future modelling. We now give an outline discussion of these.

\begin{itemize}

    \item The model atmosphere on which the present calculations are based is a semi-empirical model derived from non-LTE radiative transfer calculations in which the temperature stratification is adjusted to give good agreement with observations. For various reasons, different authors will produce different atmospheres depending on their setup and assumptions. Thus, our final results may well differ if another model atmosphere had been used, and this may affect the agreement with observations. This means that the main point of the present work is the importance of using the correct modelling and where the differences lie. For this reason, we have shown the results side-by-side between the two atomic modelling schemes. It would be interesting to know had the new atomic modelling been used in the \citet{fontenla2014} calculations whether it would have produced the same agreement with observations without the need to rely on ``missing opacity''. Alternatively, it would be interesting to know whether it would have resulted in a different temperature stratification altogether in the model atmosphere.

    \item The height variation of the micro-turbulent velocities in the \citet{fontenla2014} atmosphere were obtained from SUMER data, and are almost a constant 25\,km\,s$^{-1}$ from 10\,000\,K and upwards. These velocities are much higher than other observed velocities from neutrals and low charge states, as measured by \cite{boland1975} for example. A test calculation showed that the intensity of the hydrogen continuum was affected by altering the velocities, not just the lines. Although not the focus of this work, it was seen that the lines produced by \lw\ when radiative, thermal and micro-turbulent broadening from the \citet{fontenla2014} velocities were included were almost as broad as the SUMER resolution, which is about 0.11\,\AA\ FWHM \citep[see the discussion in][]{rao2022ntv}. Thus, different micro-turbulent velocities could be tested for their effect on the continua and lines.

    \item The huge amount of work done by atomic physicists over many years has made it possible to improve the models for the important elements here. Despite this, atomic data for neutral and low charge states is relatively scarce because of the complexity of the calculations. In addition, the models for heavier elements in the solar atmosphere (Ca, Fe, Ni) were incomplete for some of the important processes, such as CT, RR and DR. The present work has highlighted a need for such data.

    \item Following on from the previous item, it was noted from our calculations of the Si$^0$ PI cross sections that the cross sections used to produce the APAP RR data were notably smaller. In addition, the low temperature component of DR was caused by transitions to some states close to threshold, which according to experimental values from NIST \citep{nist} are not so near the threshold. Similarly, the states in the C DR calculation by \citet{altun2004,altun2004err} that contribute significantly at lower temperatures are closer in the calculation to threshold than in NIST. It is, thus, possible that the recombination from C$^+$ may be overstated, explaining why the C continua here are too low compared to observations. These issues deserve further investigation, to check how much the recombination rates would be affected and whether they alter the C and Si continua.

    \item The observations used here were averaged from the quiet Sun and yet the model atmosphere was from a cell model. As mentioned earlier, a blend of cell and network models might be required to properly reproduce observations. Similarly, there is surprisingly little observational data on UV continua, with a large scatter of values. Investigation of variations in the continua over time, as well as between cells and network, would give a better understanding of how to blend the model atmospheres and how to compare them with observations.

    \item We presented results for the solar UV wavelength range. Many cool stars have similar atmospheres as the Sun, and so the implications of the present work should extend to modelling stellar UV spectra. Investigations could also be made into any solar or stellar region where emission or absorption are in non-LTE. As highlighted in Sect.\;\ref{sec:methods}, level-resolved processes are important for low density environments in astrophysics. There could be some impact of this type of modelling in those environments because of the importance of RR into highly-excited states that may not be included in the models, as well as the potentially significant contributions from low temperature DR.

    \item The present modelling, once refined, could be used to probe the chromosphere by combining it with an advanced treatment of the physical processes occurring there, such as time dependent ionisation and 3D radiative transfer. This is particularly relevant to the recently funded Solar Atmospheric Modelling Suite (SAMS)\footnote{https://sams-project-uk.github.io/}, which will explore these effects. They can also be combined with upcoming, state-of-the-art observations from Solar-C Extreme UV Spectroscopic Telescope, for example.
 
\end{itemize}

\section{Conclusions}
\label{sec:concl}

To summarise, the updated data and modelling has had a significant impact on the ion balances and UV continua, especially for Si. Only the impact on ionisation and recombination was explored in the present work, but it was clear that level resolved photo-ionisation crossed sections are required if they are to be used for calculating the reverse, radiative recombination reaction. Otherwise, this neglects contributions to recombination from metastable levels in the recombining ion. RR rates into levels higher than those included in each charge state must be added, in order to give the correct total RR rate. Dielectronic recombination must be included where there is a significant contribution at lower temperatures. Accurate charge transfer data is also important because it made a significant contribution to the ionisation and recombination rates of Si and S.

As discussed in the previous section, there are numerous implications for this work, not least the impact it may have had on solar model atmospheres. Although only tested on quiet Sun, static conditions in one dimension, it is very likely that the modelling will be important for any other conditions on the Sun, such as prominences, active regions and flares. Given that many radiative transfer codes have similar types of atomic models, the results should impact on the present set of advanced RT codes and on modelling the atmospheres of cool stars in the UV.

\section*{Acknowledgements}

We are very grateful to Prof. Pete Storey for the support and advice received during the course of this work. We are also grateful to the referee, Prof. Petr Heinzel, for helpful comments and insights about the work. RPD and GDZ acknowledge support from STFC (UK) via the consolidated grant to the atomic astrophysics group at DAMTP, University of Cambridge (ST/T000481/1).
CMJO is grateful to the Royal Astronomical Society's Norman Lockyer fellowship, and the University of Glasgow's Lord Kelvin/Adam Smith Leadership fellowship for supporting this work.
The UK APAP data used in this work was produced over the past 20 years thanks to funding by the UK PPARC/STFC with a series of grants led by the  late Nigel R. Badnell\footnote{https://doi.org/10.3390/atoms13060055}.

\section*{Data Availability}

 The new level-resolved atomic data are made available at ZENODO (https://doi.org/10.5281/zenodo.16679094) in \lw\ format and ascii format for easy inclusion in other codes.



\bibliographystyle{mnras}
\bibliography{lw_atoms} 




\appendix

\section{Recombination data sources and suppresssion of dielectronic recombination}
\label{sec:dr_suppress}

For the new atomic models RR and DR data calculated by the following authors were incorporated: \citet{badnell2006,zatsarinny2004, abdel-naby2012, kaur2018, colgan2003, colgan2004, colgan2004err, altun2004, altun2004err, altun2006, altun2007, bautista2007, dufresne2021picrm}. 

As discussed in Sect.\;\ref{sec:drmethods}, DR suppression in the present work is estimated using the effective recombination rates of \citet{summers1972} and \citet{summers1974}. This is for the following reasons. DR suppression is less likely to take place at temperatures below approximately 10$^4$\,K. At such temperatures, radiative stabilisation of the auto-ionising state usually occurs via decay of the Rydberg electron, leaving a core-excited state that is usually close in energy to the ground \citep{storey1981}. The resulting states are unlikely to be collisionally re-ionised in these types of cases, meaning there will be little or no DR suppression. Using \citet{summers1972} will not suppress the low temperature component of the DR rates because effective rates were not produced below 10$^4$\,K. By contrast, the \citet{nikolic2018} approximation can show significant suppression at these temperatures.

\begin{figure}
\centering
\includegraphics[width=1.0\columnwidth]{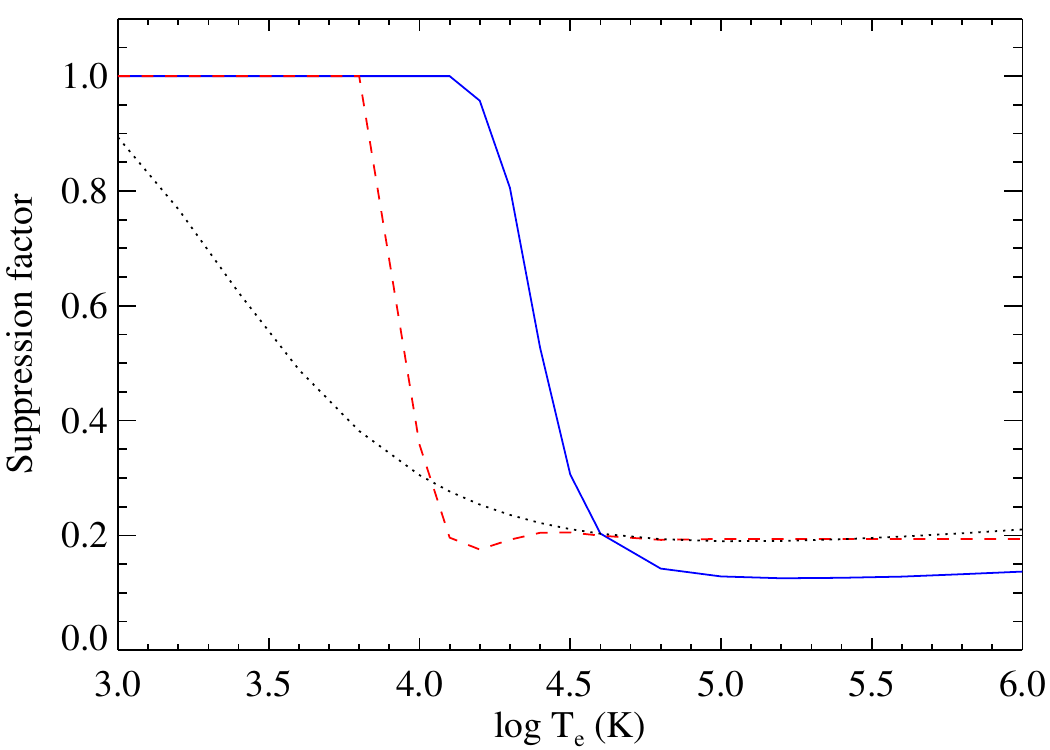}
\caption{Comparison of DR suppression factors for \ion{S}{ii} at electron density 10$^{10}$\,cm$^{-3}$ from different works: blue solid line - \citet{dufresne2021picrm}, red dashed - \citet{summers1972}, black dotted - \citet{nikolic2018}.}
\label{fig:sfactors}
\end{figure}

An example of this is illustrated here for \ion{S}{ii}. The \ion{S}{ii} DR rates published in \citet{dufresne2021picrm}, which have a significant low-temperature component below 20\,000\,K, are used in the present work. \citet{dufresne2021picrm} produced suppression factors by using those DR rates in a detailed collisional-radiative model, similar to \citet{summers1974} but using the model of \citet{storey1995}. The suppression factors are illustrated in Fig.\;\ref{fig:sfactors}, alongside those from \citet{summers1974} and \citet{nikolic2018}. According to the present models \ion{S}{ii} is populated in the temperature range 5\,000--25\,000\,K (log $T_{\rm e}$\,[K]$\approx$3.7--4.5), and here the electron densities are in the range 10$^{10}$--10$^{11}$\,cm$^{-3}$ in the current model atmosphere. The suppression factors of \citet{summers1974} follow a very similar trend to \citet{dufresne2021picrm}. They are slightly shifted in temperature because \citet{summers1972} uses the \citet{burgess1964dr} general formula, rather than \textit{ab initio} rates. By contrast, using \citet{nikolic2018} suppression factors would reduce DR rates by factors of two to five in temperatures of the solar chromosphere. It is important to highlight this because it was found for Si, as an example, that suppressing the DR rates by these amounts caused RR to be the dominant recombination mechanism over DR. This, in turn, caused the Si continuum emission to become much more enhanced than observations.

\section{Charge transfer}
\label{sec:ct_refs}

The charge transfer data were collated from calculations which used the most accurate methods for the low temperature requirements here. The sources of these data are: \cite{stancil1998c1rt}, \cite{stancil1998c1cs}, \cite{errea2015}, \cite{janev1988}, \cite{errea2000}, \cite{errea2015}, \cite{tseng1999}, \cite{liu2003} and \cite{yan2013} for C; \cite{kimura1996si1}, \cite{clarke1998},\cite{wang2006si4}, \cite{stancil1999si4he} and \cite{stancil1997si5he} for Si; and, \cite{zhao2005s1}, \cite{christensen1981}, \cite{bacchus1993s3}, \cite{zhao2005s3he}, \cite{butler1980}, \cite{stancil2001s5} and \cite{wang2002s5he} for S.


\bsp	
\label{lastpage}
\end{document}